\documentclass[11pt]{article}

\usepackage{amssymb,amsthm,amscd,array}
\usepackage{graphics}
\usepackage[final]{epsfig}	

\let\ssection=\section
\renewcommand{\section}{\setcounter{equation}{0}\ssection}

\newcommand{\bbR}{\mathbb{R}}

\newcommand{\bbT}{\mathbb{T}}

\newcommand{\Vect}{\mathrm{Vect}}
\newcommand{\Vir}{\mathrm{Vir}}

\newcommand{\ad}{\mathrm{ad}}

\newcommand{\fg}{\mathfrak{g}}

\newcommand{\fa}{\mathfrak{a}}

\chardef\s=110
\chardef\g=103

\begin{document}

\newtheorem{thm}{Theorem}[section]
\newtheorem{lem}[thm]{Lemma}
\newtheorem{cor}[thm]{Corollary}
\newtheorem{prop}[thm]{Proposition}
\newtheorem{rmk}[thm]{Remark}
\newtheorem{exe}[thm]{Example}
\newtheorem{defi}[thm]{Definition}

\def\a{\alpha}
\def\b{\beta}
\def\d{\delta}
\def\g{\gamma}
\def\om{\omega}
\def\r{\rho}
\def\s{\sigma}
\def\vfi{\varphi}
\def\vr{\varrho}
\def\l{\lambda}
\def\m{\mu}

\title{Bi-Hamiltonian nature of the equation\\
$u_{tx}=u_{xy}\,u_y-u_{yy}\,u_x$}

\author{V. Ovsienko}

\date{}

\maketitle

{\abstract{
We study non-linear integrable partial differential equations
naturally arising as bi-Hamiltonian Euler equations
related to the looped cotangent Virasoro algebra.
This infinite-dimensional Lie algebra (constructed
in \cite{OR}) is a generalization of
the classical Virasoro algebra to the case of two space
variables. Two main examples of integrable equations we obtain
are quite well known.
We show that the relation between these two equations
is similar to that between the Korteweg-de Vries and 
Camassa-Holm equations. }}

\bigskip


\bigskip


\thispagestyle{empty}

\section{Introduction}

The differential equation
\begin{equation}
\label{OEqn}
u_{tx} = u_{xy}\,u_y-u_{yy}\,u_x,
\end{equation}
where $u=u(t,x,y)$ and where $u_x,u_y$, etc. are the partial
derivatives, is a nice example of a non-linear integrable
model.  This equation is quite well known and appears
in the Mart\'{\i}nez Alonzo-Shabat ``universal hierarchy'' 
(see \cite{MS}, formula (8)).

The main purpose of this note is to show that 
this equation
(coupled together with  another differential equation,
see formula (\ref{ThmSys}) below)
naturally appears as a bi-Hamiltonian vector field
(in particular, the variable $t$ plays the
r\^ole of time while $x,y$  are space variables).
More precisely, we will show that
this equation is an Euler equation on the space dual to the
``looped cotangent Virasoro algebra'' introduced in \cite{OR}.
This, in particular, implies its integrability
in the (weak algebraic) sense of existence of a hierarchy of
first integrals in involution.

The bi-Hamiltonian approach to the same Lie algebra has
already been considered in \cite{OR} and led to another
non-linear differential equation:
$$
f_t = f_x\,\partial^{-1}_xf_y-f_y\,u
+c\,\partial^{-1}_x\,f_{yy},
$$
which can be rewritten without non-local terms:
\begin{equation}
\label{OREqn} 
u_{tx}=u_{xx}\,u_y-u_{xy}\,u_x +c\,u_{yy},
\end{equation}
after the substitution $f=u_x$.
Here $c\in\bbR$ is an arbitrary constant (the ``central charge'').
Note that this equation is also a quite well known
integrable system (see \cite{FK,FK1} and also \cite{Dun})
that appears both in differential geometry and hydrodynamic.

Equations (\ref{OREqn}) and (\ref{OEqn}) look alike
but they are not equivalent to each other.
We will show that the relation between these equations
is similar to that between the
classical Korteweg-de Vries equation (KdV) and the
Camassa-Holm equation (CH). Recall that both KdV and CH are
bi-Hamiltonian systems on the dual of the Virasoro algebra,
see \cite{Mag,Dis,KO,Mis,KM}. 
An interesting
``tri-Hamiltonian'' viewpoint was suggested in \cite{OlRo}, in
order to establish a certain duality between KdV and CH.
Equations (\ref{OREqn}) and (\ref{OEqn}) are dual in the same
sense.

This paper fits into the general framework
due to V.I. Arnold, see \cite{Arn}. 
Non-linear partial equations are viewed as
Euler equations on the dual of a Lie algebra (for instance,
the Lie algebra of vector fields). This approach explains the
geometric meaning of the equations: every Euler equation
describes geodesics of some left-invariant metric on the
corresponding group (of diffeomorphisms). 

The bi-Hamiltonian Euler equations are of special interest.
Most of the known bi-Hamiltonian non-linear partial differential equations
(KdV,CH, etc.) are of dimension
$1+1$ (i.e., contain only one space variable).
Equations (\ref{OEqn}) and (\ref{OREqn}) provide with
examples of such equations in the $(2+1)$-dimensional case.

\section{The bi-Hamiltonian formalism on
the dual of a Lie algebra}

In this section we recall the general construction of pairs of
compatible Poisson structures on the space dual to a Lie
algebra. 
We also give the standard construction of bi-Hamiltonian
vector fields on this space, due to F. Magri \cite{Mag}.

Let $\fa$ be a (finite-dimensional) Lie algebra, the
canonical
Lie-Poisson(-Berezin-Kirillov-Kostant) bracket on $\fa^*$ is
given by
\begin{equation}
\label{LieP}
\left\{
F,G
\right\}(m)=
\langle
\left[
d_mF,d_mG
\right],m
\rangle,
\end{equation}
where $m\in\fa^*$ and where $d_mF$ and $d_mG$ are the
differentials of $F$ and $G$ at $m$ understood as elements of
$\fa$, namely $dF_m\in(\fa^*)^*\cong\fa$.
This Poisson structure is linear, i.e.,
the space of linear functions equipped with the bracket
(\ref{LieP}) is a Lie subalgebra of $C^\infty(\fa^*)$ (isomorphic to
$\fa$).

Given a skew-symmetric bilinear form $\om:\fa\wedge\fa\to\bbR$,
one defines another Poisson structure on $\fa$:
\begin{equation}
\label{ConPS}
\left\{
F,G
\right\}_\om(m)=
\om
\left(
d_mF,d_mG
\right).
\end{equation}
This structure is with constant coefficients, i.e.,
the bracket of two linear functions is a constant function
on $\fa^*$.

Two Poisson structures are called
compatible (or a Poisson pair) if their
linear combination is again a Poisson structure.
The following simple fact is well known 
(see, e.g., \cite{AG}, Section 5.2).

\begin{prop}
\label{WKProp}
The Poisson
structures (\ref{LieP}) and (\ref{ConPS}) are compatible
if an only if $\om$ is a 2-cocycle on $\fa$.
\end{prop}

The simplest example of a constant Poisson structure
(\ref{ConPS}) corresponds to the case where the 2-cocycle $\om$
is trivial (i.e., a coboundary).
Every such structure is of the following form.
Fix a point $m_0\in\fa^*$ and set
\begin{equation}
\label{CoBord}
\om(x,y)=
\left\langle
m_0,\,[x,y]
\right\rangle.
\end{equation}
It worth noticing that one can understand this particular case
of constant Poisson structure on $\fa^*$ as
the most general one.
Indeed, it suffices to replace $\fa$ by its central
extension.

Every function $H$ on $\fa^*$ defines two vector fields
that we denote $X_H$ and $X^\om_H$ on $\fa^*$:
the first one is Hamiltonian with respect to the
linear structure (\ref{LieP}) and is given by
\begin{equation}
\label{GETypeEq}
X_H(m)=\ad^*_{d_mH}\,m,
\end{equation}
while the vector field $X^\om_H$ is Hamiltonian 
with respect to the constant bracket (\ref{ConPS}).
In the particular case (\ref{CoBord}), one has explicitly
\begin{equation}
\label{CstEq}
X^\om_H(m)=\ad^*_{d_mH}\,m_0.
\end{equation}

Given two compatible Poisson structures,
a vector field which is Hamiltonian with respect to 
the both structures is called bi-Hamiltonian.
The usual way to construct bi-Hamiltonian
vector fields on $\fa^*$ is as follows. 
Consider the following 1-parameter family of Poisson
structures
$$
\{\,,\,\}_{\l}=\{\,,\,\}_\om-\l\,\{\,,\,\},
$$
(parameterized by $\l\in\bbR$).
Assume that $H$ is a Casimir function of this bracket, 
i.e., one has
$$
\{H,\,F\}_{\l}=0,
\qquad
\hbox{for all}
\quad F\in{}C^\infty(\fa^*).
$$
Assume also that $H$ is written in a form of a series
\begin{equation}
\label{Her}
H=H_0+\l\,H_1+\l^2\,H_2+\cdots
\end{equation}
One immediately obtains the following facts:

\begin{enumerate}
\item
the function $H_0$ is a Casimir function of $\{\,,\,\}_\om$;
\item
the Hamiltonian vector field corresponding to $H_k$
are bi-Hamiltonian, namely
$$
X_{H_k}=
X^\om_{H_{k+1}},
$$
for all $k$;
\item
all the functions $H_k$ are in involution with respect to 
the both Poisson structures, indeed, for $k\leq\ell$ one has
$$
\{H_k,\,H_\ell\}=
\{H_{k+1},\,H_\ell\}_\om=
\{H_{k+1},\,H_{\ell-1}\}=\cdots=0,
$$
and therefore are first integrals of every vector field $X_{H_k}$.
\end{enumerate}

Let us summarize the method.
To construct an integrable hierarchy,
one chooses a function $H_0$ which is a Casimir function of
the constant Poisson structure $\{\,,\,\}_\om$;
one then considers its Hamiltonian vector
field, $X_{H_0}$, with respect to the Lie-Poisson structure.
This vector field is again Hamiltonian with respect to 
the constant Poisson structure, with some Hamiltonian function
$H_1$, so that one has:
$X_{H_0}=X^\om_{H_1}$.
One then iterates the procedure
to find $H_2,H_3$, etc.

\section{The looped cotangent Virasoro algebra and its dual}

In this section we recall the definition \cite{OR} of the
looped cotangent Virasoro algebra. 
We also describe its coadjoint representation.

Let us start with the definition of the classical Virasoro
algebra. 
Consider the Lie algebra, $\Vect(S^1)$, of vector
fields on the circle:
$
f(x)\,\frac{\partial}{\partial x}
$
where $f\in{}C^\infty(S^1)$ and $x$ is a coordinate on $S^1$,
we assume $x\sim{}x+2\pi$.
To simplify the formul{\ae}, we will identify
$\Vect(S^1)$ with $C^\infty(S^1)$;
the Lie bracket in $\Vect(S^1)$ is then given by
$$
\left[
f,g
\right]=
f\,g_x-f_x\,g.
$$
The Virasoro algebra, $\Vir$, is a (unique up to isomorphism) 
one-dimensional central extension of $\Vect(S^1)$.
It is defined on the space
$\Vect(S^1)\oplus\bbR$, the commutator being given by
\begin{equation}
\label{Vir}
\left[
(f,\,\a),\,(g,\,\b)
\right]=
\left(f\,g_x-f_x\,g,\,
\int_{S^1}f\,g_{xxx}\,dx
\right).
\end{equation}
Note that the constants $\a$ and $\b$ do not enter the right
hand side of the above formula since they belong to the
center of $\Vir$.

The Virasoro algebra was found by Gelfand and Fuchs
\cite{GF}, the constant term in the right hand side of
(\ref{Vir}) is called the Gelfand-Fuchs cocycle.
This Lie algebra plays an important r\^ole
in mathematical physics, essentially because of the
applications of its representations to conformal field theory,
but also because of its applications to integrable systems.

The dual space, $\Vect(S^1)^*$, is the space of distributions.
One often considers only a subspace,
$\Vect(S^1)^*_\mathrm{reg}$, called the ``regular dual'' (cf.
\cite{Kir}). As a vector space, this regular dual is, again,
isomorphic to $C^\infty(S^1)$, the pairing
$\langle.,.\rangle:\Vect(S^1)\otimes{}C^\infty(S^1)\to\bbR$
being give by
$$
\left\langle
f(x)\,\frac{\partial}{\partial x},\,a(x)
\right\rangle:=
\int_{S^1}f(x)\,a(x)\,dx.
$$
The regular dual to the Virasoro algebra is
$\Vir^*_\mathrm{reg}=C^\infty(S^1)\oplus\bbR$;
the coadjoint action of $\Vir$ on its regular
dual is:
$$
\ad^*_{(f,\a)}(a,\,c)=
\left(
f\,a_x+2\,f_x\,a+c\,f_{xxx},\,0
\right).
$$
This formula easily follows from (\ref{Vir}) and the definition of
$\ad^*$, see \cite{Kir}.
Note that the constant $c$ is preserved by the action,
it is therefore a parameter  called the central
charge.

\begin{rmk}
{\rm
The Virasoro algebra is, indeed, exceptional.
The reason is that the Lie algebras of vector fields on
a manifold of dimension $\geq2$, has no central extensions,
cf. \cite{Fuk}.
The problem of generalization of the Virasoro algebra
is an interesting subject studied by many authors.
}
\end{rmk}

The looped cotangent Virasoro algebra \cite{OR} is a
generalization of $\Vir$ in the case of two variables.
We consider the 2-torus $\bbT^2$ and define a Lie algebra
structure on the space
$$
\fg=
{C^\infty(\bbT^2)}
\oplus{C^\infty(\bbT^2)}
\oplus\bbR^2.
$$
given by the commutator
\begin{equation}
\label{CommutHatEq}
\left[
\left(
\begin{array}{l}
\displaystyle
f\\[10pt]
\displaystyle
a\\[10pt]
(\a,\a')
\end{array}
\right),\,
\left(
\begin{array}{l}
\displaystyle
g\\[10pt]
\displaystyle
b\\[10pt]
(\b,\b')
\end{array}
\right)
\right]
=
\displaystyle
\left(
\begin{array}{l}
\displaystyle
f\,g_x-f_x\,g\\[6pt]
\displaystyle
f\,b_x+2\,f_x\,b-g\,a_x-2\,g_x\,a\\[6pt]
\displaystyle
\Big(
\int_{S^1\times{}S^1}f\,g_{xxx}\,dxdy\,,\,
\int_{S^1\times{}S^1}
\left(
f\,b_{y}-g\,a_y
\right)\,dxdy
\Big)
\end{array}
\right)
\end{equation}
where $(x,y)$ are the usual
coordinates on
$\bbT^2$ and where $f,g,a,b$ are smooth functions in $x,y$;
the constants $\a,\a',\b,\b'\in\bbR$ are elements of the
center.
Note that, unlike the Virasoro algebra,
the center of $\fg$ is two-dimensional.

\begin{rmk}
{\rm
One notices that the quotient-algebra $\fg/\bbR^2$ (by the
center) is the loop algebra with coefficients in the
semidirect sum
$\Vect(S^1)\ltimes\Vect(S^1)^*_\mathrm{reg}$.
The dependence in $y$-variable in this quotient-algebra is somehow
trivial.
The second 2-cocycle in (\ref{CommutHatEq}), however,
makes this dependence in $y$ non-trivial.
Note also that this cocycle is
rather similar to the Kac-Moody cocycle.
}
\end{rmk}

We will need the coadjoint representation of $\fg$
and the notion of regular dual space.
Consider the pairing
$\langle.,.\rangle:\fg\otimes\fg\to\bbR$
$$
\left\langle
\left(
\begin{array}{l}
\displaystyle
f\\[4pt]
\displaystyle
a\\[4pt]
\displaystyle
(\a_1,\a_2)
\end{array}
\right),\,
\left(
\begin{array}{l}
\displaystyle
g\\[4pt]
\displaystyle
b\\[4pt]
\displaystyle
(\a_1,\a_2)
\end{array}
\right)
\right\rangle=
\int_{S^1\times{}S^1}
\left(
f\,b+g\,a
\right)dxdy
+\a_1\b_1+\a_2\b_2,
$$
that identifies $\fg$ with a part of its dual space:
$\fg\hookrightarrow\fg^*$, we call this subspace
the regular dual space of $\fg$ and denote it by
$\fg^*_{\mathrm{reg}}$.
The coadjoint action of $\fg$ on $\fg^*_{\mathrm{reg}}$
can be easily calculated:
\begin{equation}
\label{CExtnEFor}
\widehat{\ad}^*_{\left(
\begin{array}{l}
f\\
a\\
(\a_1,\a_2)
\end{array}
\right)
}\,
\left(
\begin{array}{l}
\displaystyle
g\\[4pt]
\displaystyle
b\\[4pt]
(c_1,c_2)
\end{array}
\right)
=
\left(
\begin{array}{l}
\displaystyle
f\,g_x-f_x\,g+c_2\,f_y\\[4pt]
\displaystyle
f\,b_x+2\,f_x\,b-a_x\,g-2\,a\,g_x+c_1\,f_{xxx}+c_2\,a_y\\[4pt]
(0,0)
\end{array}
\right).
\end{equation}
Note that the center $\bbR^2\subset\fg$ acts trivially.

The Lie algebra $\fg$ is infinite-dimensional.
In order to define the brackets
(\ref{LieP}) and (\ref{ConPS}) in this case,
we consider only the space of so-called
pseudodiffe\-rential polynomials
on $\fg^*_{\mathrm{reg}}$:
$$
H(f,\,a)=
\int_{S^1\times{}S^1}
h\left(
f,\,a,
\,f_x,\,a_x,\,f_y,\,a_y,
\,\partial^{-1}_xf,\,\partial^{-1}_xa,
\,\partial^{-1}_yf,\,\partial^{-1}_ya,\,f_{xy},\,a_{xy},\ldots
\right)
\,dxdy,
$$
where $h$ is a polynomial and $f,a,
f_x,a_x,f_y,a_y,
\partial^{-1}_xf,\ldots$ are understood as independent
variables. 

The differential $d_mH$ is replaced by the
standard variational derivative:
$$
d_{(f,\,a)}H:=
\left(
\d_a{}H,\,\d_fH
\right)
$$
understood as element of $\fg/\bbR^2$.
The Lie-Poisson structure (\ref{LieP}) then makes sense on
$\fg^*_{\mathrm{reg}}$ and the Hamiltonian vector fields are
again given by (\ref{GETypeEq}). 

\begin{exe}
{\rm
Recall that the Euler-Lagrange equation provides an explicit
formula for variational derivatives.
For instance, one has
$$
\begin{array}{rcl}
\displaystyle
\d_aH&=&
\displaystyle
h_a-
\partial_x
\left(
h_{a_x}
\right)-
\partial_y
\left(
h_{a_y}
\right)-
\partial^{-1}_x
\left(
h_{\partial^{-1}_xa}
\right)
-
\partial^{-1}_y
\left(
h_{\partial^{-1}_ya}
\right)\\[14pt]
&&
\displaystyle
+(\partial_x)^2
\left(
h_{a_{xx}}
\right)
+\partial_x\partial_y
\left(
h_{a_{xy}}
\right)
+(\partial_y)^2
\left(
h_{a_{yy}}
\right)
\pm\cdots
\end{array}
$$
where, as usual, $h_u$ means the partial derivative
$\frac{\partial{}h}{\partial{}a}$, similarly
$h_{a_x}=\frac{\partial{}h}{\partial{}a_x}$, etc..
}
\end{exe}

One of course should be careful with the definition
of the non-local operators $\partial^{-1}_x$ and
$\partial^{-1}_y$.
We use the expression
$$
(\partial^{-1}_xf)(x,y)=
\int_{0}^{x}f(\xi,y)\,d\xi-
\int_{0}^{2\pi}f(x,y)\,dx,
$$
and similarly for $\partial^{-1}_y$.

We refer to \cite{Dis} for
further details on Hamiltonian formalism on
infinite-dimensional (functional) Lie algebras.

\subsection{Calculating the bi-Hamiltonian equations}

Let us fix the following point of $\fg^*_\mathrm{reg}$:
\begin{equation}
\label{PointEq}
m_0=
\left(
f(x),\,a(x),\,c_1,\,c_2
\right)_0=
\left(
1,\,1,\,0,\,c
\right),
\end{equation}
with arbitrary $c\in\bbR$, and consider the constant Poisson
structure (\ref{ConPS}) corresponding to the coboundary
(\ref{CoBord}).
The Hamiltonian vector field $X^\om_H$ with the Hamiltonian $H$
is then given by
$$
\begin{array}{rcl}
\displaystyle
f_t &=&
-\left(
\d_aH
\right)_x+
c\left(
\d_aH
\right)_y\\[8pt]
\displaystyle
a_t &=&
2\left(
\d_aH
\right)_x-
\left(
\d_fH
\right)_x+
c\left(
\d_fH
\right)_y.
\end{array}
$$
The limit case $c\to\infty$ corresponds to the following
structure
\begin{equation}
\label{PointVFEq}
\begin{array}{rcl}
\displaystyle
f_t &=&
\left(
\d_aH
\right)_y\\[8pt]
\displaystyle
a_t &=&
\left(
\d_fH
\right)_y.
\end{array}
\end{equation}

We are ready to formulate our main result.

\begin{thm}
\label{TheThm}
The following system on $\fg^*_\mathrm{reg}$
\begin{equation}
\label{ThmSys}
\begin{array}{rcl}
\displaystyle
u_{tx} &=&
\displaystyle
u_{xy}\,u_y-u_{yy}\,u_x
\\[10pt]
\displaystyle
v_{tx}  &=&
\displaystyle
2\left(
u_{yy}\,v_x-u_{xy}\,v_y
\right)
+u_{y}\,v_{xy}-u_{x}\,v_{yy}
\\[6pt]
&&
\displaystyle
-2\left(
u_{yy}\,u_x
+2\,u_{xy}\,u_y
\right)
\end{array}
\end{equation}
is bi-Hamiltonian with respect to the standard Lie-Poisson
structure on $\fg_\mathrm{reg}$, together with
(\ref{PointVFEq}), where $f=u_y$ and $a=v_y$.
\end{thm}

\begin{proof}
The simplest class of Casimir functions of this constant
Poisson structure are linear combinations of the functionals
$\int{}f\,dxdy$ and $\int{}u\,dxdy$.
We will choose the Casimir function
$$
H_0(f,\,a)=
\int_{S^1\times{}S^1}
\left(a-f\right)\,dxdy.
$$
The Hamiltonian vector field, $X_{H_0}$,
with respect to the Lie-Poisson structure defines the following
vector field
\begin{equation}
\label{PerEq}
\begin{array}{rcl}
\displaystyle
f_t &=&
f_x\\[6pt]
\displaystyle
a_t &=&
2\,f_x+a_x.
\end{array}
\end{equation}
Indeed, one obviously has
$\left(\d_a{H_0},\,\d_f{H_0}\right)=
\left(-1,\,1\right)$ (understood as an element of $\fg/\bbR^2$)
and one then applies the definition (\ref{GETypeEq}).

One thus looks for a function $H_1(f,\,a)$ on 
$\fg^*_\mathrm{reg}$ such that
its Hamiltonian vector field with respect to the
constant Poisson structure satisfies 
$$
X^\om_{H_1}=X_{H_0},
$$
which leads to the following
system of equation on the variational derivatives $\d_f{H_1}$
and
$\d_u{H_1}$:
$$
\begin{array}{rcl}
\displaystyle
-\left(
\d_a{H_1}
\right)_x+
c\left(
\d_a{H_1}
\right)_y &=&
f_x\\[8pt]
\displaystyle
2\left(
\d_a{H_1}
\right)_x-
\left(
\d_f{H_1}
\right)_x+
c\left(
\d_f{H_1}
\right)_y &=&
2\,f_x+a_x.
\end{array}
$$
Introducing the first-order differential operator
$$
\Lambda=-\partial_x+c\,\partial_y,
$$
one shows by a simple straightforward calculation that
following function:
\begin{equation}
\label{H1}
H_1(f,\,a)=
\int_{S^1\times{}S^1}
\left(
\Lambda^{-1}(f_x)\,a+
\Lambda^{-1}(f_x)\,f-
\Lambda^{-2}(f_{xx})\,f
\right)dxdy
\end{equation}
is a solution of the above system.

The Hamiltonian vector field $X_{H_1}$ is then as follows
$$
\begin{array}{rcl}
\displaystyle
f_t &=&
\displaystyle
\Lambda^{-1}(f_x)\,f_x-
\Lambda^{-1}(f_{xx})\,f
+c_2\,\Lambda^{-1}(f_{xy})\\[10pt]
\displaystyle
a_t &=&
\displaystyle
\Lambda^{-1}(f_x)\,a_x+2\,\Lambda^{-1}(f_{xx})\,a
-\Lambda^{-1}(a_{xx})\,f-2\,\Lambda^{-1}(a_x)\,f_x\\[6pt]
&&
\displaystyle
-2\left(
\Lambda^{-1}(f_{xx})-\Lambda^{-2}(f_{xxx})
\right)f
-4\left(
\Lambda^{-1}(f_{x})-\Lambda^{-2}(f_{xx})
\right)f_x
\\[6pt]
&&
\displaystyle
c_1\,\Lambda^{-1}(f_{xxxx})
+c_2\left(
\Lambda^{-1}(a_{xy})+2\,\Lambda^{-1}(f_{xy})
-2\,\Lambda^{-2}(f_{xxy})
\right)
\end{array}
$$
In the same way as in \cite{Mis}, we substitute to this
equation
$
f=\Lambda(u)
$
and
$
a=\Lambda(v)
$
and rewrite it in the following
form:
\begin{equation}
\label{CH1}
\begin{array}{rcl}
\displaystyle
-u_{tx}+c\,u_{ty} &=&
\displaystyle
c
\left(
u_{xy}\,u_x-u_{xx}\,u_y
\right)
+c_2\,u_{xy}\\[10pt]
\displaystyle
-v_{tx}+c\,v_{ty}  &=&
\displaystyle
c
\left(
2\,u_{xx}\,v_y-2\,u_{xy}\,v_x
+u_{x}\,v_{xy}-u_{y}\,v_{xx}
\right)
\\[6pt]
&&
\displaystyle
2\left(
u_{xx}-\Lambda^{-1}(u_{xxx})
\right)
\left(
u_x-c\,u_y
\right)
+4\left(
u_{x}-\Lambda^{-1}(u_{xx})
\right)
\left(
u_{xx}-c\,u_{xy}
\right)
\\[6pt]
&&
\displaystyle
c_1\,u_{xxxx}
+c_2\left(
v_{xy}+2\,u_{xy}
-2\,\Lambda^{-1}(u_{xxy})
\right).
\end{array}
\end{equation}
It is very easy to check that, in the limit case $c\to\infty$,
this system coincides with (\ref{ThmSys}) with exchanged
notation for the variables $(x,y)\leftrightarrow(y,x)$.
\end{proof}

Theorem \ref{TheThm} implies the existence of an
infinite series of first integrals in involution for the
equation (\ref{OEqn}), as well as
of an infinite hierarchy of commuting flows, see \cite{OR},
Section 5.5.

\begin{rmk}
{\rm
1)
The special case $c=0$ in (\ref{PointEq}) was considered
in the details in \cite{OR}.
This case is related to the equation (\ref{OREqn}).

2)
One can also choose a non-zero value of the first
central charge $c_1$ in (\ref{PointEq}).
This will, however, only change the second equation in
(\ref{CH1}). 

3)
Consider the first equation in (\ref{CH1}).
The term $c_2\,u_{xy}$ can be removed by the
transformation
$u\mapsto{}u-\frac{c_2}{c}\,x$.
Furthermore, the coordinate transformation
$(x,y)\to(x,y+c\,x)$ leads to the following family:
$$
u_{tx}=c\left(
u_{xy}\,u_y-u_{yy}\,u_x
\right)
+u_{xx}\,u_y-u_{xy}\,u_x
$$
depending on $c$ as parameter.
This family gives one an interpolation
between the equations (\ref{OEqn}) and (\ref{OREqn}),
but with zero central charge.
}
\end{rmk}

\vskip 0.5cm

\textbf{Acknowledgments}.
I am grateful to G. Misiolek, E. Ferapontov, A. Reiman and C.
Roger for enlightening discussions.
A part of this work was done during the meeting
``M\'ecanique g\'eom\'etrique'', November 2007 at CIRM;
I am pleased to thank the organizers A. Constantin and B.
Kolev.

\vskip 0.5cm


Institut Camille Jordan

Universit\'e Claude Bernard Lyon 1,

21 Avenue Claude Bernard,

69622 Villeurbanne Cedex,

FRANCE;

ovsienko@math.univ-lyon1.fr

\end{document}